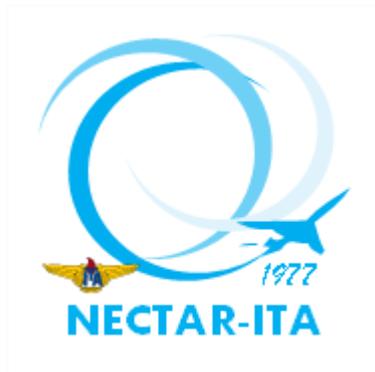

# DOCUMENTO DE TRABALHO

An empirical analysis of the determinants of network construction for Azul Airlines

Bruno Felipe de Oliveira

Alessandro V. M. Oliveira

**Instituto Tecnológico de Aeronáutica**
**São José dos Campos, Brasil, 2022**

# An empirical analysis of the determinants of network construction for Azul Airlines

Bruno Felipe de Oliveira✈
Alessandro V. M. Oliveira

This version: November 15, 2021.

## Abstract

This paper describes an econometric model of the Brazilian domestic carrier Azul Airlines' network construction. We employed a discrete-choice framework of airline route entry to examine the effects of the merger of another regional carrier, Trip Airlines, with Azul in 2012, especially on its entry decisions. We contrasted the estimated entry determinants before and after the merger with the benchmarks of the US carriers JetBlue Airways and Southwest Airlines obtained from the literature, and proposed a methodology for comparing different airline entry patterns by utilizing the kappa statistic for interrater agreement. Our empirical results indicate a statistically significant agreement between raters of Azul and JetBlue, but not Southwest, and only for entries on previously existing routes during the pre-merger period. The results suggest that post-merger, Azul has adopted a more idiosyncratic entry pattern, focusing on the regional flights segment to conquer many monopoly positions across the country, and strengthening its profitability without compromising its distinguished expansion pace in the industry.

______________________________

✈ Corresponding author. Email address: brunofo@ita.br.

▪ Affiliations: Center for Airline Economics, Aeronautics Institute of Technology, Brazil.

▪ Acknowledgements: The first author wishes to thank the Coordination for the Improvement of Higher Education Personnel (CAPES) - Finance Code 001. The second author wishes to thank the São Paulo Research Foundation (FAPESP) - grant n. 2020-06851; National Council for Scientific and Technological Development (CNPq) - grant n. 301344/2017-5. The authors wish to thank the anonymous reviewers, Marcelo Xavier Guterres, Mauro Caetano de Souza, Victor Rafael Rezende Celestino. All remaining errors are ours.

# 1. Introduction

Route entries, their effects, and what drives carriers to announce new destinations to their networks are essential topics that still hold interest in the air transport industry. Market entry, in general, is a highly explored empirical theme in many areas such as economics, management, and marketing literature (Dixit & Chintagunta, 2007), the reason being that it not only spurs competition but may induce innovation (Hüschelrath & Müller, 2011). Not surprisingly, one of the crucial elements of both incumbent and newcomer airlines' strategic planning that can be impacted by the competitive dynamics that follow market entry is their business model as transformations and adaptations to newly established market environments may emerge.

This study considers the case of Azul Airlines in the Brazilian air transport industry. Azul Airlines was established in 2008 by David Neeleman, founder of JetBlue, one of the largest low-cost carriers (LCCs) in the US market, and certainly a good reference for the newcomer's startup period. From serving only three destinations in December 2008, Azul has become the third largest airline in Brazil, and now serves more domestic destinations than any other carrier since the new millennium.[1] Not even the giant legacy carrier Varig and its group of regional carriers in the early 2000s had a network that reached the mark of 110 served destinations with scheduled flights in Brazil.

Azul typically explores short- and medium-haul markets from smaller airports, a characteristic that, in principle, reminds us of Southwest Airlines (Boguslaski et al., 2004). It also operates the only major secondary airport in the country, São Paulo/Campinas (VCP). However, contrary to the notable US LCC, Azul is far from maintaining a standardized fleet, and since its launch has entered a diversified portfolio of low-, medium-, and high-density routes. So far, the only constraint to the carrier's expansion has probably been the lack of availability of departure and landing slots at São Paulo/Congonhas Airport (CGH), a key airport in Brazil.

One motivation for studying the case of Azul Airlines is to investigate its change in its business model across the years and how these changes may have affected its network and route entry decisions. A major landmark in Azul's journey was its merger with Trip Airlines in May 2012, when it was just three years old. On the one hand, it still resembled a low-cost carrier, with a clear penetration pricing strategy to quickly promote traffic growth and consumer awareness in more densely populated markets such as São Paulo, Porto Alegre, Salvador, Rio de Janeiro, and Belo Horizonte metropolitan regions. Trip, on the other hand, was the largest regional carrier in Latin America with an extensive network coverage across the country.

---

[1] "Azul S.A. Form 20-F." Filed with the *United States Securities and Exchange Commission*, April 30, 2021, available at ri.voeazul.com.br. Figures relative to 2020.

In this study, we developed an econometric model of network construction for Azul Airlines in the Brazilian airline industry. Our aim was to identify the effects of the 2012 merger and how this event affected the carrier's route entries. So far, the literature has not discussed the possible effects of a merger on an airline's entry decision and how it might affect its network planning. This study fills this gap by investigating how the merger with Trip impacted Azul's entry decision in the post-merger period. We employed a discrete-choice framework of entry decisions, applying the simple probit, random-effects probit, and rare event logistic regression models to a data set that comprised more than one million observations from thousands of domestic airport pairs across the 2008–2018 period.

We contrasted the estimated entry determinants before and after the merger with the benchmarks of JetBlue Airways and Southwest Airlines obtained from the literature, namely Boguslaski, Ito, and Lee (2004), and Müller, Hüschelrath & Bilotkach (2012), hereafter BIL04 and MHB12, respectively. Additionally, we propose a methodology for comparing different airline entry patterns by utilizing the Kappa statistic for inter-rater agreement (Cohen, 1960).

The next sections of this research are divided as follows: Section 2 presents the literature review on entry in airline markets, and the studies that covered the empirical issue of airline entry patterns. Section 3 presents the research design, with a description of the application, the data, the econometric model, and the estimation strategy. Section 4 presents the estimation results and our proposed method for entry patterns comparisons. Finally, Section 5 concludes.

## 2. Literature review

The worldwide expansion of the airline industry is closely tied to the Airline Deregulation Act in the US and the subsequent growth of low-cost carriers. Understanding how airlines plan their network and market entries would inevitably lead to the way to understand how LCCs work.

The original business model of LCCs had the following characteristics.[2] They served short-haul routes, used regional or secondary airports, operated point-to-point, and had limited (or without) customer loyalty programs, limited passenger services (no frills), a high proportion of bookings made through the internet, high fleet utilization, and a standardized fleet.

Today, most LCCs do not strictly follow the expected characteristics of their business models. Due to the dynamic nature of the civil aviation industry, airlines are becoming more standardized in terms of their operations and business models to achieve higher operational and financial efficiency. This standardization in the airline industry is also known as business model convergence or hybridization (Daft and Albers, 2015; Jarach et al., 2009; Urban et al., 2018). Despite the airline hybridization trend, many studies in the literature still adopt the classification of low-cost carriers and full-service carriers to refer to certain airlines, depending on their initial business model. This study also uses these classifications to refer to certain airlines as LCCs, such as Southwest, JetBlue, and Azul, despite the gradual change in their original business model over the years.

---

[2] IATA Economics Briefing n.5, 2006.

## *2.1. Entry into airline markets*

Early research on the impacts of an LCC entry assessed how these airlines could affect airfares and flight demand in the airline industry. With the success of Southwest Airlines, the largest LCC in the world, this business model has been replicated worldwide; the model strategy has even changed, moving towards hybridization. At the same time, new airlines have entered the industry with a new way of applying low-cost philosophy, such as ultra LCCs (ULCCs).

Although change in pricing behavior was one of the first studied effects of low-cost entry, the topic remains a favorite with researchers. The research is motivated by changes in the air travel market, such as the hybridization of airline business models and the entry of ULCCs. Studies show that LCCs can still reduce airfares (Asahi & Murakami, 2017; Chen, 2017; Zhang et al., 2018; Ren, 2020), but they are no longer as effective when compared to ULCCs (Bachwich & Wittman, 2017; Zou et al., 2017). In addition to affecting rival pricing behaviors, many studies found, LCCs could affect the operation and business strategy of their rivals. Additionally, they can affect the capacity decisions of other airlines, such as the size of the aircraft used on a particular route or frequency of flights, and even force rivals to change their flight times to avoid competition (Pearson et al., 2015; Sun, 2015; Bendinelli et al., 2016; Mohammadian et al., 2019). Current studies on charter flights have also confirmed the trends in the literature, as low-cost airlines have effectively replaced charter airlines (Wu, 2016; Castillo-Manzano et al., 2017). Overall, studies show that low-cost carriers force their rivals to respond to them to not lose their dominant position in the market, by reducing their airfares or adapting their operations for better efficiency.

Several studies have also investigated the effect of LCCs on airport revenue by examining whether airports were experiencing any financial benefits from an association with an LCC but no clear consensus has emerged. While some studies have shown a negative or zero effect of LCC on financial efficiency (Yokomi et al., 2017; Zuidberg, 2017), others have reached the opposite conclusion (Augustyniak et al., 2015; Button et al., 2017; Martini et al., 2020). Given this lack of consensus, Tavalaei and Santalo (2019) argue that, to assess the real effects of LCCs on airports' financial performance, researchers should consider the strategic purity of the airport or the number of LCCs operating there. As airport managers increasingly struggle to obtain competitive advantages, especially in areas with multiple airports, there have been several studies that have concluded that airport managers should offer LCC flights at their airports as a way to increase their connectivity (Zhang et al., 2017) and even offer facilitation services for passengers who opt for self-connections (Chang et al., 2019). The latter segment of the industry has not yet been widely explored by airports but has potential for future growth (Cattaneo et al., 2017)

There are also studies on the general effects of LCCs on air travel demand. Recent studies still find the Southwest effect, which is the rising passenger demand on routes served by LCCs (Rolim et al., 2016; Boonekamp et al., 2018). However, some of these studies have problems with positive marginal effects due to the limited number of LCC flights analyzed in different regions (Valdes, 2015; Tsui & Fung, 2016). Recent studies have confirmed the vast literature on tourism demand that shows the positive effects of LCCs on tourism (Alsumairi & Tsui, 2017). Moreover, some suggest that the airport or even the local government should attempt to attract LCCs to bolster the local economy (Álvarez-Díaz et al., 2019). Studies aver that LCCs can promote the development of a region, even when it is not a tourist destination (Bowen Jr., 2016; Taumoepeau et al., 2017).

When analyzing the effects of LCC on passenger choice, other studies have shown that the presence of an LCC is an essential factor for air transport systems users to exercise their choices. Most studies have explored general passenger and airline choice decisions (Kim, 2015; Saffarzadeh et al., 2016; Paliska et al., 2016; Yang, 2016; Hunt & Truong, 2019), while some have analyzed specific air transport system users, such as students checking the availability of a low-cost route between their cities and the university (Cattaneo et al., 2016) or tourists choosing a destination served by LCC flights (Clavé et al., 2015). Borhan et al. (2017) even expanded the literature by investigating how LCCs can influence automotive drivers to change transport modes, showing that low-cost flights can be an essential factor for drivers to start flying.

*2.2. Airline entry patterns*

The literature on airline network construction is generally tied to the US airline market deregulation in 1978. Most of the research tend to analyze the airline market as a whole, not focusing on LCCs. For example, Morrison and Winston (1990) analyzed the dynamics of airline pricing and competition in the airline industry. Regarding the entry and exit of airlines, they used a probit model and found that when a carrier is already operating at a pair of airports, it significantly impacts its entry decision owing to the knowledge of the demand, and a competitor's activity at a pair of airports does not discourage entry. According to them, a high fare on a given route negatively affects entry decisions. Although this was not an intuitive and expected result, the authors (1990) explained that some of these high-fare routes presented entry barriers, high costs, or incumbent carriers' aggressive responses, affecting the estimation of coefficients.

Joskow et al. (1994) found a similar result to Morrison and Winston (1990) as entry is driven by cost factors, specifically that airlines tend to enter city pairs if the price (or fare) is low. Sinclair (1995) further expanded the literature by showing strong evidence that airlines' entry and exit decisions are affected by the size and utilization of a hub-and-spoke system; an incumbent with a robust hub system can inhibit entry, while an entrant with a robust hub system will enter the market.

Dresner et al. (2002) studied the effects of barriers on entry decisions, showing that slot controls, gate constraints, and gate utilization during peak hours negatively affect entry decision. Gil-Moltó and Piga (2008) analyzed the European airline market in terms of low-cost and traditional carrier entry. Among the different variables tested, some of them confirmed the existing literature results, like the previous presence in a city pair, but some variables presented interesting results. For example, the number of companies already operating on a route is positively correlated with entry. The authors explained this result as a lower number of companies in a route is due to the presence of a dominant airline or entry barriers. Alternatively, the market size presented a negative correlation, which can be explained by the dominant airline or entry barrier. American and European airlines have a hub model in which some airlines dominate certain airports using them as operating hubs. Owing to their operations' scale, even with a large market size, it acts as an entry barrier and inhibits competitors' entry.

Regarding the literature on LCC entry patterns, Ito and Lee (2003) analyzed LCCs' growth in the US airline industry and factors that influence their entry. According to them, the most important predictor of an LCC entry is market density. In their study, the price variable positively affected entry decision, contrary to previous research on network carriers, for example, Morrison & Winston (1990) and Joskow et al. (1993), showing that LCCs concentrate their entry in markets where incumbents were earning a large price markup. Boguslaski et al. (2004) further expanded the literature by analyzing Southwest's entry strategy evolution over the years, finding a change in behavior in choosing routes to operate, from dense and short to thin and long-haul markets. Both these works suggest that LCCs are no longer bound to fly only dense and short-haul markets and serve leisure passengers, and that network carriers face increased exposure to LCC competition over time.

Warnock-Smith and Potter (2005) used a qualitative approach to this matter and concluded that high demand is an essential factor for LCCs to choose which airports to enter. According to them, other essential factors are quick turnaround facilities, convenient slot times, and high airport competition. Analyzing the LCC entry patterns in Brazil, Oliveira (2008) concluded that airline entry behavior was consistent with the classic Southwest entry pattern, focusing on dense and short-haul routes. He also found exciting evidence that the Brazilian LCC had changed its entry pattern following its foundation, pairing with JetBlue's entry pattern and focusing on long-haul routes, and explained it as an effect of the country's idiosyncrasies, such as unobserved economies of scope.

Müller et al. (2012) studied the entry pattern of LCC JetBlue Airlines in the US domestic airline industry. They showed that JetBlue consistently avoided concentrated airports and instead targeted concentrated routes by using secondary airports on thicker routes, avoiding competition with network carriers. They also showed that JetBlue targeted longer-haul markets on non-stop markets and avoided slot-restricted airports and routes already operated by other LCCs.

Boguslaski, Ito, and Lee (2004) and Oliveira (2008) conclusion that LCC entry patterns changed over time was also studied by de Wit and Zuidberg (2012) in their work on the growth limits of the LCC model. They analyzed the European and American airline markets and concluded that the continental market showed signs of saturation for LCCs. They identified new business strategies adopted by LCCs, including shifting to primary airports, hubbing, entering codeshare agreements or alliances, and acquiring or merging with other airlines.

Homsombat et al. (2014) studied the route entry strategy of Qantas Airways and its low-cost subsidiary Jetstar Airways in the Australian domestic market. They concluded that Jetstar tended to enter routes where other LCCs compete directly with them, and the group's main market strength was its presence in a large number of routes, both from full-service carriers (Qantas) and LCCs (Jetstar). Zhang et al. (2017) examined the Australian airline industry by focusing on regional segments. In line with Homsombat et al. (2014), their results suggest a strategic competition between airline brands. Wang et al. (2020) explored the airline market in New Zealand, suggesting favorable regional socioeconomic and tourism factors as relevant determinants of route entry.

In the Asian market, Fu et al. (2015) investigated the case of Spring Airlines, an LCC in China, one of the largest emerging air markets but with strict civil aviation regulations. The Chinese government controls fuel supply, airport fares, aircraft purchases, and even entry decisions as routes connecting to its main airports (Fu et al., 2014; 2015). Even so, these regulations have not prevented LCCs from operating in this market. According to Fu et al. (2015), Spring Airlines' strategy is to enter routes with high average fare prices, and allowing a high aircraft load factor by offering cheaper tickets. Another strategy adopted by Spring is to enter routes not linked to major airports in the country (partly because of regulations), except its hub Shanghai, which suggests that Spring prefers to link its new destinations with Shanghai airport.

Wang et al. (2017) continued this discussion and analyzed the entry patterns of LCCs in the Hong Kong market, which is regulated by the Chinese government. The authors found results consistent with those found by Fu et al. (2015), and pointed out that low-cost airlines in this market prefer dense markets with high per capita income, and that the presence of other airlines on the route did not discourage entry.

Dobruszkes et al. (2017) studied the LCC shift to primary airports, one of the new business strategies cited by Wit and Zuidberg (2012). They also analyzed the European and US low-cost airline markets, reasoning that both markets are maturer than other regions. Their work analyzed data from Ryanair and Southwest until 2015 and confirmed the trend of increased operations from major airports, which implies an increase in competition between LCCs and traditional network carriers.

Nguyen and Nguyen (2018) studied the entry and exit patterns of US airlines. According to their model, seven significant factors explain entry decisions: total number of passengers, average market fare, number of carriers, distance, presence of an LCC, origin hub, and destination hub. Exit decisions too can be explained by all of these factors, with the addition of route type and the business model for the dominant airline on the route.

Atallah et al. (2018) analyzed the evolution of airline strategies in the US market using data from 2005 to 2015, before and after the recession, to investigate any change in these companies' entry patterns. The authors found results consistent with previous research on low-cost airlines in terms of their entry into the country's primary airports and highly competitive routes. The authors also found a strategic difference between LCCs and FSCs: while the former focused on entering different markets in this period, testing new routes for unmet demand, the latter focused on offering higher flight frequency on existing routes, trying to strengthen its position in the market. Similarly, Zou and Yu (2020) analyzed the evolution of entry patterns from Southwest and JetBlue using data from 1993 to 2016. Their results show that both airlines tend to enter routes that can be the dominant carriers. They also confirmed the results by Ito and Lee (2003), showing that Southwest has indeed changed its entry behavior over time, entering longer and less dense routes. The results also confirmed the findings of Müller et al. (2012), showing that JetBlue consistently prefers longer distance routes. Finally, their results show that, at some point, the presence of an LCC in the route was a deterrent to both Southwest and JetBlue, but over the years, its significance has been decreasing.

The literature on airline entries has been concerned with analyzing LCCs' changing strategies in mature markets, such as the US and Europe. Additionally, the literature has also analyzed these airlines' behavior in emerging markets, showing that despite being studied since US civil aviation deregulation, it is still a relevant topic in the industry. Understanding where the literature will go from this point on is essential to understand where the airline market itself is going. In the literature, authors such as de Wit and Zuidberg (2012) pointed out new business strategies that LCCs could adopt, including a shift to primary airports, hubbing, codeshare agreements, alliances, and mergers and acquisitions. While Dobruszkes et al. (2017) have already studied the shift to the primary airport, and hub airports have been analyzed since Sinclair (1995), the other two strategies have not yet been well explored in the literature and could be further investigated.

Many other studies have recently investigated the impact of entry in airline markets, such as Zhang et al. (2017), Fu et al. (2019), and Zhang et al.. (2019), Valido et al. (2020), and Wang et al. (2020). Wang et al. (2017) estimate that Asian LCCs prefer large markets with large populations and high incomes and traffic volumes. Gaggero and Piazza (2021) show that airlines tend to follow the market leader and enter a route served by the incumbents, and that entry is more likely when the airline operates other routes at the two endpoint airports of a route. Zou and Yu (2020) investigated the similarity and dissimilarity between Southwest and JetBlue, and found that the former has become more attracted to routes with connections to be better positioned for the business passenger segment. Bet (2021) shows evidence that airline incumbents' actions are more prone to deter, rather than accommodate, entry.

## 3. Research design

### *3.1. Application*

We investigated some of the main factors that may have driven Azul Airlines' route entry and exit decisions in the domestic market since the late 2000s. Neeleman is a serial entrepreneur of the airline industry, who founded Morris Air, WestJet, JetBlue Airways, and Breeze Airways. He was motivated to invest in the Brazilian airline industry at that time by fast market growth and high concentration. In December 2008, Azul started its hub operations based on São Paulo/Campinas Airport (VCP), a secondary airport in the São Paulo metroplex region. From only two nonstop destinations out of VCP at its launch, Azul has considerably expanded its network to become the third largest airline, with more than 110 destinations across the country, 168 aircraft and 27.7% market share.[3] The two largest airlines in the domestic market are the low-cost carrier Gol Airlines (38.1%) and the full-service carrier LATAM Airlines (33.7%). VCP has considerably benefited from Azul's growth, from less than a million passengers in 2008 to more than 10 million six years later. Azul's other hubs are Belo Horizonte/Confins Airport (CFN) and Recife Airport (REC).

At the time of their merger in 2012,[4] Trip was a Brazilian and then largest Latin American regional airline. With this merger, the Azul-Trip group became the third largest airline in Brazil, serving 96 destinations. Post-merger, there was a period in which the Azul-Trip group adjusted its network by exiting non-profitable cities until reaching the current network configuration, serving 464 city-pairs as of December 2018.

---

[3] "Azul S.A. Form 20-F." Filed with the *United States Securities and Exchange Commission*, April 30, 2021, available at ri.voeazul.com.br. Figures relative to 2020.

[4] "Brazilian carriers Azul and Trip unveil merger plans," *Flight Global*, May 28, 2012, available at www.flightglobal.com.

**Figure 1 - Azul's network evolution from 2008 to 2021**

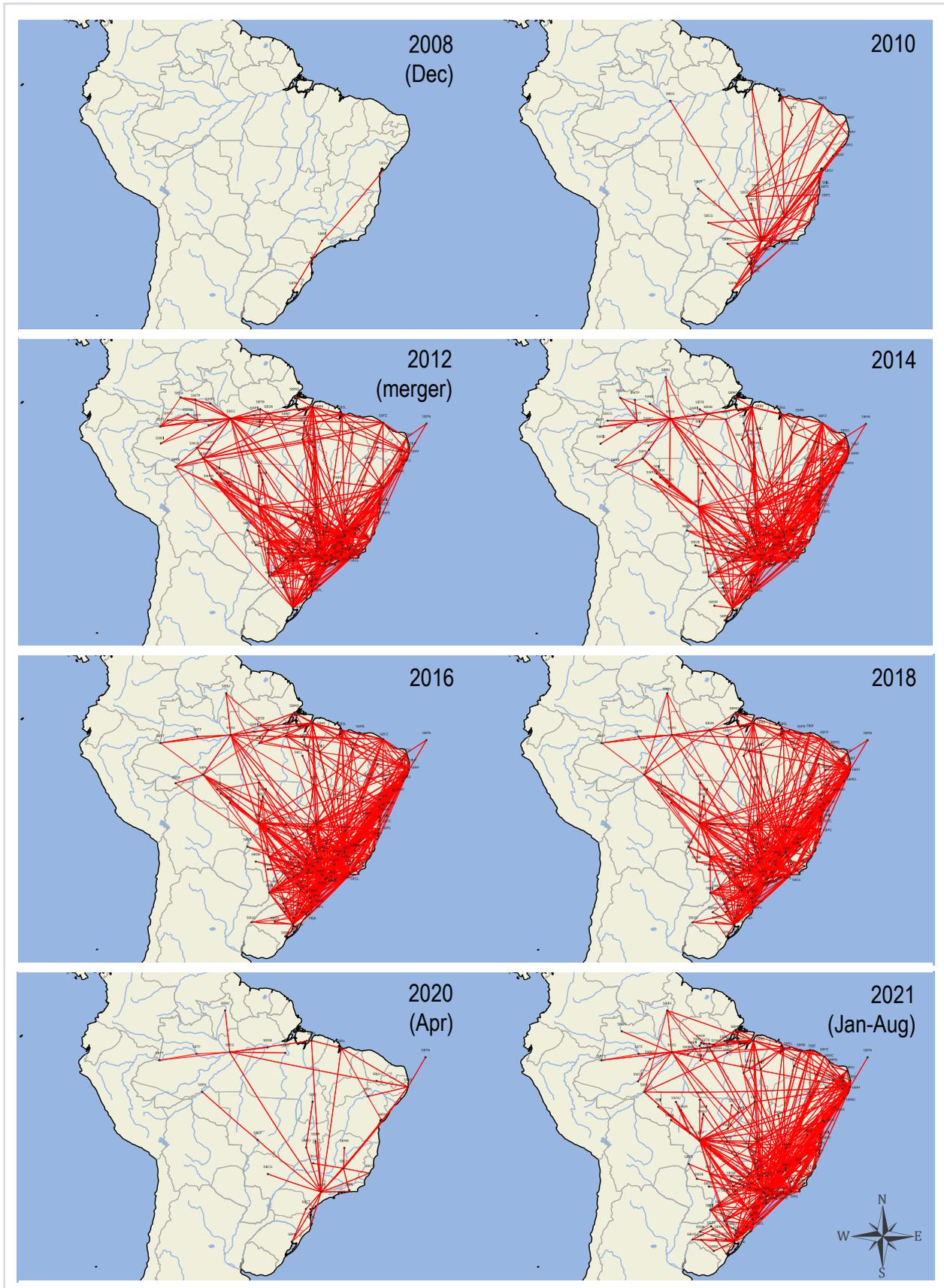

*Source: National Civil Aviation Agency's Air Transport Statistical Database (aggregate), with own computations.*

After the 2012 merger, Azul presented a notable evolution in the domestic air travel market. The carrier had a 14.2% growth in the number of enplanements between 2013 and 2018, against its peers' 1.4% growth. In addition, it reported the largest earnings before interest and taxes (EBIT) in the industry in five out of the six years from 2013 to 2018.[5]

Figure 1 displays the notable evolution of Azul's domestic network since 2008 until mid-2021. Although the COVID-19 pandemic significantly affected its operations in 2020, Azul has quickly recovered from the crisis, becoming the first airline in Brazil to reach pre-pandemic levels in 2021.[6]

### 3.2. Data

We collected passenger and traffic data for this study from the National Civil Aviation Agency (ANAC)'s Air Transport Statistical Database (aggregate), which comprises air transport supply and demand information aggregated at the route level (airport-pair/airline/month). Our definition of route is the directional airport pair. Our dataset comprises a panel of domestic routes in Brazil, with yearly observations between 2008 (the year of Azul's first flight) and 2018.

We designed a procedure for considering an amplified set of possible airport pairs of interest for Azul, aiming at eliminating the possible selection bias that may emerge from our sample definition choices. To create the dataset, we first identified all airports used by domestic commercial aviation (scheduled and non-scheduled) between January 2000 and August 2021. We considered a time span that is longer than the final sample's period length aiming at incorporating all possible airports of interest to Azul. In all, 312 airports were identified, which made it possible to create 97,032 possible pairs of airports. Of these, 6,581 airport pairs (6.8%) had at least one flight in the period. As in BIL04, flights with distances below 100 miles and above 3,000 miles were discarded (a total of 14,674 discards, or 1,334 airport pairs). The final balanced panel data set contains 1,582,140 observations (97,032 routes times 11 years, minus 14,674 discards).

In our dataset, we focused on the Azul-Trip merger event, which took place in 2012. To simplify the presentation of the empirical results, we denote the complete sample by "FULL". We then consider two temporal sub-samples: the period before the Azul-Trip merger, denoted by "BEF" (2008-2011), containing 382,792 observations (95,698 routes times 4 years), and the period after the merger, denoted by "AFT" (2012-2018), containing 669,886 observations (95,698 routes times 7 years). Additionally, we used the year 2007, prior to Azul's entry into the industry, as a base case for differentiating between existing and the new routes, denoted as "EXIST" and "NEW", respectively. We classified all airport pairs that had operations in 2007 as existing and the remaining routes as new.

---
[5] Sources: ANAC's Air Transport Demand and Supply Report and Financial Statements of Brazilian Airlines (2013-2018).

[6] "Azul é a primeira companhia a recuperar índices pré-pandemia, mostram dados da Anac", *O Globo*, September 24, 2021, available at oglobo.globo.com.

Most data are publicly available from the National Civil Aviation Agency (ANAC). Other sources of information are the Brazilian Institute of Geography and Statistics (IBGE), and the Central Bank of Brazil. To compute the socioeconomic measures related to the endpoint airports we considered the concept of "mesoregions", i.e., groupings of nearby cities. All monetary variables were adjusted by a deflator based on the Extended National Consumer Price Index (IPCA) of IBGE.

### 3.3. Econometric model

We developed an empirical model of network construction of Azul Airlines in the domestic airline industry in Brazil. We built our model upon BIL04 and MHB12, with most variables being proxies for the variables employed in those studies. Although one of our goals was to maximize the comparability of our results with the findings of this previous literature, we also introduced a set of covariates not utilized before.

Equation (1) presents the specification of our proposed empirical model.

$$
\begin{aligned}
AZ^*_{k,t} = &\ \beta_1 PAX_{k,t} + \sum_i \beta_{2,i} DIST\ i_k + \beta_3 POP_{k,t} + \beta_4 INC_{k,t} + \beta_5 UNEMPL_{k,t} + \\
& \beta_6 VACATION_{k,t} + \beta_7 SECND_k + \beta_8 SLOT_k + \beta_9 FEE_{k,t} + \beta_{10} NETWEC_{k,t} + \\
& \beta_{11} ZERAZCIT_{k,t} + \beta_{12} AZSHCON_{k,t} + \beta_{13} HUBOTH_{k,t} + \beta_{14} NONHUB_{k,t} + \\
& \beta_{15} HHI_{k,t} + \beta_{16} MAXHHI_{k,t} + \beta_{17} MAXHHI_{k,t} \times NONHUB_{k,t} + \beta_{18} FSCMAJ_{k,t} + \\
& \beta_{19} LCCMAJ_{k,t} + \beta_{20} LCCCOMP_{k,t} + \beta_{21} BANKR_{k,t} + \beta_{22} REGSMA_{k,t} + \\
& \beta_{23} NEW_{k,t} + \beta_{24} TREND_t + \beta_{25} TREND_t \times DIST_k + \beta_{26} TREND_t \times HUB_k + \\
& \beta_{27} TREND_t \times SECND_k + \beta_{28} TREND_t \times NEW_{k,t} + \varepsilon_{k,t},\ AZ_{k,t} = \mathbb{1}[AZ^*_{k,t} > 0],
\end{aligned}
\qquad(1)
$$

where $AZ^*_{k,t}$ denotes a latent variable expressing Azul's network decisions regarding entry in airport-pair $k$ at period $t$. $AZ_{k,t}$ is a binary response variable accounting for Azul's airport-pair entry, the function $\mathbb{1}[.]$ is the indicator function for the binary outcome, meaning that entry occurs ($AZ_{k,t} = 1$) when $AZ^*_{k,t} > 0$. $AZ_{k,t}$ is assigned with value one only for the year the carrier starts flight operations on the route, being equal to zero for the other years. $\varepsilon_{k,t}$ is the error term, and the $\beta$s are the unknown parameters. In Table 1, we provide a presentation of the regressors present in Equation (1). Detailed information on each variable is available in the Appendix. We also provide correlation coefficients in the Appendix (Table 5). To simplify the exposition, henceforth we omit indexes $k$ and $t$.

**Table 1 - Description of model variables**[7]

| Regressor | Description | Level/Computation | Metric | BIL04 equivalent | MHB12 equivalent |
|---|---|---|---|---|---|
| PAX | revenue passengers (in 2007) | airport-pair | count (ln) | dense | Passengers |
| DIST, DIST2 | distance, distance squared | airport-pair | miles | − | Distance, Distance$^2$ |
| DIST X | distance greater than X miles | airport-pair | mut. exclv. dummies | D(distanceX) | − |
| POP | population | geom. mean (O, D cities) | count (ln) | meanpop | Population |
| INC | GDP per capita | geom. mean (O, D cities) | BRL deflated (ln) | − | Income |
| UNEMPL | unemployment rate proxy | geom. mean (O, D cities) | index [2004=100] | − | Unempl. |
| VACATION | tourism revenues over GDP | geom. mean (O, D cities) | proportion | max(vacation) | − |
| SECND | secondary airport | maximum (O, D airports) | dummy | − | Secondary airp. |
| SLOT | slot airport | maximum (O, D airports) | dummy | − | Slot restr. airp. |
| FEE | airport landing fees | geom. mean (O, D cities) | BRL deflated (ln) | − | PFC |
| NETWEC | number of cities served by Azul | sum (O, D cities) | count | − | Netw. economies |
| MAXAZCIT | number of cities served by Azul | maximum (O, D airports) | year mean | max(swcities) | − |
| MINAZCIT | number of cities served by Azul | minimum (O, D airports) | year mean | min(swcities | − |
| ZERAZCIT | endpoint city not served by Azul | maximum (O, D airports) | dummy | D(swzero) | − |
| AZSHCON | Azul's share of connecting pax | airport-pair | proportion | swshare | − |
| HUB | endpoint airport is Azul hub | maximum (O, D airports) | dummy | − | − |
| HUBOTH | endpoint airport is Azul's rival hub | maximum (O, D airports) | dummy | D(hub) | − |
| NONHUB | endpoint airport is not a major hub | maximum (O, D airports) | dummy | − | Non-HUB |
| HHI | route concentration (in 2007) | airport-pair | index [0,1] | markethhi | Route HHI |
| MAXHHI | airport concentration (in 2007) | maximum (O, D airports) | index [0,1] | max(cityhhi) | Airp. HHI |
| MINHHI | airport concentration (in 2007) | minimum (O, D airports) | index [0,1] | min(cityhhi) | − |
| LCCCOMP | presence of LCCs (in 2007) | airport-pair | dummy | D(lowcost) | LCC comp. |
| BANKR | presence of bankrupt | airport-pair | dummy | − | Chapter 11 route |
| MININC | GDP per capita | minimum (O, D cities) | BRL deflated (ln) | min(income) | − |
| MAXINC | GDP per capita | maximum (O, D cities) | BRL deflated (ln) | max(income) | − |
| EXIST | existing route (with respect to 2007) | airport-pair | dummy | − | Existing market |
| NEW | new route (with respect to 2007) | airport-pair | dummy | − | − |
| FSCMAJ | presence of major FSC (in 2007) | airport-pair | dummy | − | − |
| LCCMAJ | presence of major LCC (in 2007) | airport-pair | dummy | − | − |
| REGSMA | presence of small regional (in 2007) | airport-pair | dummy | − | − |
| TREND | time trend | systemwide | discrete sequence | − | − |

---

[7] "BIL04" denotes Boguslaski, Ito & Lee (2004); "MHB12" denotes Müller, Hüschelrath & Bilotkach (2012). Note that we omit subscripts $k$ and $t$. See details of each variable in the Appendix.

Note that in Table 1, we indicate the equivalence of each variable with respect to BIL04 and MHB12. For example, in Table 1 we indicate that POP is a covariate in our framework that accounts for the size of the population at each endpoint airports' cities of the route. The equivalent variables in BIL04 and MHB12 are "meanpop" and "Population", respectively.[8] Also note that in Table 1 more variables are listed than those present in Equation (1). This is due to the fact that, later in 4.2, we will use the full specifications of BIL04 and MHB12, which involve some additional variables, such as DIST, DIST2, MAXAZCIT and MINAZCIT.

It is important to emphasize, however, that one of the limitations of our approach is the fact that BIL04 adopts the city pair concept, unlike MHB12. As the model of Equation (1) assumes the route definition as being the airport pair, it is possible that our results are more consistent with MHB12 than with BIL04. We aimed to compensate for this possible problem by using a proportion of regressors slightly more in line with the BIL04 specification.

### *3.4. Estimation strategy*

To estimate the unknown parameters of Equation (1), we utilized a probit as the baseline estimator, denoted as "PROBIT". To challenge the results produced by the probit estimator, we also employed the Random-Effects Probit Model ("XTPROBIT"), and the Rare Event Logistic Regression model ("RELOGIT"). Whereas the first is a probit-based model suitable for panel data (Guilkey & Murphy, 1993), the second is a logit-based model suitable when the sample is very unbalanced, with one outcome being rarer than the other (King and Zeng, 2001). In our case, we had panel data and a rare event of AZ = 1. In all procedures, the standard errors of estimates allowed for intragroup correlation (clustered sandwich estimator), using airport pairs as clusters.

A crucial problem in our empirical framework is the endogeneity of regressors referring to strategic industry movements after the entry of Azul Airlines. In particular, the variables PAX, HHI, MAXHHI, FSCMAJ, LCCMAJ, LCCCOMP, and REGSMA are the most susceptible to the problem of endogeneity bias. To get around this problem, we configured these variables with values from the year 2007. As discussed earlier, we use 2007 because it is a year when Azul was not present in the industry. Therefore, it is not contaminated by the possible endogenous industry movements in response to Azul's entries.

---

[8] See BIL04, page 332, and MHB, page 498.

# 4. Results and discussion

## 4.1. Estimation results

Table 2 presents the estimation results of our empirical model of the route entry determinants of Azul Airlines. Columns (1), (2), and (3) of Table 2 display the results of the estimations using the full sample period, denoted by AZ(FULL). The difference between columns refers to the estimation procedure, PROBIT, XTPROBIT, and RELOGIT. The majority of estimates in these columns maintain the same sign and statistical significance, suggesting the robustness of the results across the different estimators. For example, PAX is positive and statistically significant for the five columns. The same is true for the DIST dummies, VACATION, SLOT, and SECND, among several other covariates.

Columns (4) and (5) of Table 2 present the results of the estimates obtained by splitting the sample into two sub-samples, considering the periods before and after the 2012 Azul-Trip merger. These sub-samples are denoted as AZ (BEF) and AZ (AFT), respectively. Taken together, these results reveal crucial strategic movements concerning the carrier's network decisions and adaptations to the Brazilian market during the 2010s.

Regarding distance, Table 2 shows all the coefficients of the DIST dummies (DIST 300 - DIST 1500 flip sign from Columns (4) to (5) to be negative and statistically significant. As the base case of the dummies is constituted by markets with distances less than 300 miles, we can infer that, after the merger, Azul started to give preference to short-haul routes, more in line with the operation of a regional airline. Additionally, the coefficient of POP is positive and statistically significant only in the period before the merger, but not after the merger. This result reinforces the results of the DIST dummies, indicating that the airline started to enter less populated cities across the country. These results are consistent with the airline's fleet planning after the merger, with the 70-seater turboprop ATR-72 models, suitable for shorter flight stages owing to their lower speed, constituting approximately 30 percent of its fleet; these are also adequate for medium-density markets because of their lower seating capacity. Therefore, we have evidence suggesting that the airline has adapted its original business model to incorporate the regional segment as one of its main markets served in the industry. With the revamped strategic orientation, the carrier was able to conquer many monopoly positions in air travel markets across the country, without compromising its notable expansion pace and distinctive profitability.

**Table 2 - Estimation results: Azul entries - full model specification**

|  | (1) AZ(FULL) | (2) AZ(FULL) | (3) AZ(FULL) | (4) AZ(BEF) | (5) AZ(AFT) |
|---|---|---|---|---|---|
| PAX | 0.0934*** | 0.0945*** | 0.2930*** | 0.0474** | 0.1014*** |
| DIST 300 | -0.1786*** | -0.1848*** | -0.3906*** | 0.0233 | -0.2259*** |
| DIST 600 | -0.3738*** | -0.3838*** | -0.8889*** | 0.4185*** | -0.5398*** |
| DIST 900 | -0.4125*** | -0.4260*** | -1.0392*** | 0.2951 | -0.6269*** |
| DIST 1200 | -0.4409*** | -0.4585*** | -1.1080*** | 0.7085*** | -0.7612*** |
| DIST 1500 | -0.7161*** | -0.7401*** | -1.8576*** | 0.0068 | -1.0773*** |
| POP | -0.0360 | -0.0362 | -0.1077 | 0.4606*** | -0.0298 |
| INC | -0.0154 | -0.0179 | -0.0148 | -0.2736 | -0.0253 |
| UNEMPL | 0.0009 | 0.0006 | 0.0018 | 0.0125 | 0.0010 |
| VACATION | 1.2371*** | 1.2503*** | 2.7882*** | -1.7459* | 1.1472*** |
| SECND | 0.9649*** | 0.9995*** | 2.3349*** | 1.2303*** | -0.6633** |
| SLOT | -0.0879* | -0.0918* | -0.1974* | -0.7273*** | -0.0678 |
| FEE | -0.0048 | -0.0035 | -0.1118** | 0.4046*** | -0.0317 |
| NETWEC | 0.0336*** | 0.0346*** | 0.0694*** | 0.0475*** | 0.0287*** |
| ZERAZCIT | -0.5684*** | -0.5812*** | -1.6383*** | -1.3323*** | -0.1128*** |
| AZSHCON | 0.6092*** | 0.6217*** | 1.0738*** | 0.7072*** | 0.6104*** |
| HUBOTH | 0.0752 | 0.0729 | 0.3517*** | 0.0925 | 0.1818*** |
| NONHUB | -0.3501*** | -0.3565*** | -0.8179*** | -0.3735 | -0.4049*** |
| HHI | 0.3015*** | 0.3069*** | 0.7669*** | 0.2118* | 0.3327*** |
| MAXHHI | -0.0437 | -0.0452 | 0.0046 | -0.2748 | -0.0273 |
| MAXHHI × NONHUB | 0.2863*** | 0.2949*** | 0.6618*** | 0.1873 | 0.2849*** |
| FSCMAJ | 0.1489** | 0.1610** | 0.2242 | 0.4918*** | 0.0981 |
| LCCMAJ | -0.4761*** | -0.4913*** | -1.1601*** | -0.4239*** | -0.5416*** |
| LCCCOMP | 0.4955*** | 0.5208*** | 1.0141*** | 0.4204*** | 0.5321*** |
| BANKR | -0.3038 | -0.2986 | -0.4231 | – | -0.0982 |
| REGSMA | 0.2404*** | 0.2638*** | 0.3576*** | 0.2704*** | 0.2534*** |
| NEW | 1.2376*** | 1.2766*** | 3.2526*** | 1.1347*** | 1.5299*** |
| TREND | -0.0836*** | -0.0877*** | -0.2335*** | -0.0008 | -0.1544*** |
| TREND × DIST | 0.0002 | 0.0002 | 0.0027 | -0.0326*** | 0.0036*** |
| TREND × HUB | -0.0920*** | -0.0951*** | -0.1989*** | 0.0930*** | -0.0788*** |
| TREND × SECND | -0.2718*** | -0.2798*** | -0.5968*** | -0.6366*** | -0.0488 |
| TREND × NEW | 0.1032*** | 0.1058*** | 0.2296*** | 0.0111 | 0.0785*** |
| Estimator | PROBIT | XTPROBIT | RELOGIT | PROBIT | PROBIT |
| Airport-Pair Clusters | 95,698 | 95,698 | 95,698 | 95,698 | 95,698 |
| Log likelihood Statistic | -5,992 | -5,989 | -5,546 | -585 | -5,102 |
| Pseudo R2 Statistic | 0.5601 | 0.5363 | 0.5929 | 0.6502 | 0.5611 |
| AIC Statistic | 12,050 | 12,046 | 11,160 | 1,235 | 10,270 |
| BIC Statistic | 12,442 | 12,450 | 11,563 | 1,582 | 10,646 |
| Nr Observations | 1,052,678 | 1,052,678 | 1,052,678 | 382,792 | 669,886 |

*Notes: Estimation results in Columns (1), (4), and (5) produced by the probit model, denoted as "PROBIT"; estimation results in Column (2) produced by the random effects probit model, denoted as "XTPROBIT"; and estimation results in (6) produced by the logistic regression in rare events data of King and Zeng (2001), denoted as "RELOGIT". Standard errors of estimates allow for intragroup correlation (clustered sandwich estimator), using airport pairs as clusters. "–" denotes that the variable is dropped. "AZ" denotes the probability of Azul's route entry. "FULL", "BEF", and "AFT" denote full sample period, before merger period, and after merger period, respectively. P-value representations: ***p<0.01, ** p<0.05, * p<0.10.*

Another variable that flips sign from Column (4) to Column (5) of Table 2, is SECND. This result indicates that since the merger, the airline has placed less emphasis on new destinations from its main hub, the VCP, in a revamped strategy of greater spatial diversification. Indeed, the airline has initiated operations at new hubs, namely Belo Horizonte/Confins and Recife Airports. However, the result of the TREND × HUB variable points to the fact that the airline has started to enter markets where none of the endpoint airports are one of its hubs.

With respect to HHI, we note that since the merger, the airline has entered many highly concentrated routes. The result for this variable also suggests a preference for operating routes with fewer carriers (Column 5). The result of the MAXHHI × NONHUB variable, which becomes a positive and statistically significant coefficient after the merger (Column 5), is also consistent with this result, given that the airline has entered many non-hub airports. Likewise, the sign flipping of the ZERAZCIT variable from Columns (4) to (5) suggests a preference for entering new markets. The result of the TREND × NEW variable also points to a growing preference for entering new markets since the merger.

Finally, concerning the competition-driven network decisions, the results of the variables FSCMAJ and LCCMAJ in Column (4) reveal a preference for the carrier of entering markets already operated by the major full-service carrier LATAM and a strategy of avoiding markets operated by the major low-cost carrier Gol. However, the results point to a statistical significance of the first effect only in the period before the merger. Additionally, flight operations by medium-sized low-cost carriers and small regional carriers have the effect, ceteris paribus, of attracting Azul's entry in both periods, as seen by the positive and statistically significant coefficients of LCCCOMP and REGSMA in Columns (4) and (5).

We performed equality tests to check whether the coefficients between columns were significantly different. At the 95% confidence level, the tests rejected the null hypothesis of equal coefficients in all cases. Moreover, the differences between the coefficients of both columns indicate that the estimates are attenuated in three quarters of the cases, with 42% of the coefficients flipping their sign. In sum, the estimation results of Columns (4) and (5) reveal that Azul engaged in very different entry pattern strategies after the 2012 merger event.

*4.2. Entry patterns comparisons*

Once we estimate the empirical results of Equation (1), we investigate whether Azul's network growth observed in the sample period is consistent with the low-cost carrier benchmarks estimated by the previous studies. Therefore, we focus on comparing the estimates in Table 2 with the estimates by MHB12 and BIL04, who study the cases of JetBlue Airways and Southwest Airlines, respectively. Our ex-ante expectation is that there may be some consistency between Azul entry patterns and JetBlue's, given that they were both founded by Neeleman.

To carry out a comparative study of Azul's network decisions with US carriers, we proposed the following three-step methodology:

I. Classify each variable originally estimated by MHB12 and BIL04 as "Significant Negative", "Not Significant", and "Significant Positive", according to the sign and statistical significance of its estimated coefficient. Gather all the classifications of each model to form unique raters computed for each benchmark airline: the "original BIL04 estimates for Southwest" and the "original MHB12 estimates for JetBlue" raters.

II. Estimate two modified versions of Equation (1) employing specifications that are strictly based on MHB12 and BIL04, i.e., using as many as the same variables as possible. Again, classify each variable as "Significant Negative", "Not Significant", and "Significant Positive", and gather all the classifications to form unique raters computed for Azul: the "BIL04-like estimates for Azul" and the "MHB12-like estimates for Azul" raters.

III. Using the unique raters computed in Steps I and II, calculate the Kappa Statistic for Interrater Agreement (Cohen, 1960) to inspect the consistency of Azul's entry decisions with the entries of the benchmark airlines.

Cohen's Kappa statistic is a measure of agreement between raters. In our case, we linked each model estimation set to a corresponding rater, by classifying each coefficient into three categories: "Significant Negative", "Not Significant", and "Significant Positive" (Steps I and II). After developing the raters, we computed the Kappa statistic to inspect the consistency between them and examine their agreement regarding the entry patterns of Azul and the selected US carriers.

The Kappa statistic method for two unique raters requires pairwise comparisons of their classifications and computing the frequency of agreements between them. The statistic provides a synthesis of the comparisons, corrected for the chance agreement case. The statistic ranges from –1 to +1, with 1 being the case of perfect agreement, zero representing the case where the agreement is the same as the one expected by chance, and values lower than zero for where agreement is even weaker than expected by chance. Thus, the higher the value of kappa, the stronger is the agreement. Landis and Koch (1977, p. 165) propose the following scale for analyzing the agreement strength as assessed by the Kappa statistic: below 0.0 (poor), 0.00–0.20 (slight), 0.21–0.40 (fair), 0.41–0.60 (moderate), 0.61–0.80 (substantial), and 0.81–1.00 (almost perfect). For example, the estimated coefficients of Columns (4) and (5) reveal that the two corresponding raters agree in 38.7% of the cases, that is, 12 out of 31 coefficients. The agreement expected by chance is 33.0%, which is close to the observed interrater agreement. The computed Kappa statistic in this case is 0.0854 (standard error 0.1205), which indicates only "slight" agreement, which is not statistically significant at 5% levels.

Our interpretation of the Kappa application to the comparison of airline entry pattern estimates between Azul and the chosen benchmark carriers is as follows. If the entry patterns between Azul and each of the benchmarks are independent, then the Kappa statistic is near zero. In this case, the hypothesis of the rater agreement by chance was not rejected. If there is some consistency between the network decisions of the carriers under comparison due to, say, business model similarities, then the strength of the interrater agreement is relatively high and the chance hypothesis is rejected.

Accomplishing Step I of our proposed methodology was relatively easy. The corresponding estimates are readily accessible in each of the considered empirical studies, namely BIL04 and MHB12. As stated by Step I, we labeled this information as the "original BIL04 estimates for Southwest" and "original MHB12 estimates for JetBlue".

For Step II, however, we needed to develop our own estimates to obtain the "BIL04-like estimates for Azul" and "MHB12-like estimates Azul." Therefore, we ran alternative versions of Equation (1) in which we aimed to specify the empirical models to mimic the original specifications of MHB12 and BIL04 and apply them to our dataset. Equations (2) and (3) present the proposed specifications. All regressors contained in Equations (2) and (3) are briefly described in Table 1 and in more detail in the Appendix. First, in Equation (2), we present the model specification based on MHB12.[9]

$$\begin{aligned}
AZ_{k,t}^{*,m} = &\; \delta_1 \text{DIST}_{k,t} + \delta_2 \text{DIST SQ}_{k,t} + \delta_3 \text{PAX}_{k,t} + \delta_4 \text{HHI}_{k,t} + \delta_5 \text{LCCCOMP}_{k,t} + \\
& \delta_6 \text{BANKR}_{k,t} + \delta_7 \text{NETWEC}_{k,t} + \delta_8 \text{EXIST}_{k,t} + \delta_9 \text{SECND}_k + \delta_{10} \text{SLOT}_k + \\
& \delta_{11} \text{MAXHHI}_{k,t} + \delta_{12} \text{NONHUB}_{k,t} + \delta_{13} \text{MAXHHI}_{k,t} \times \text{NONHUB}_{k,t} + \delta_{14} \text{FEE}_{k,t} + \\
& \delta_{15} \text{POP}_{k,t} + \delta_{16} \text{INC}_{k,t} + \delta_{17} \text{UNEMPL}_{k,t} + \varepsilon_{k,t}^m, \; AZ_{k,t} = \mathbb{1}[AZ_{k,t}^{*,m} > 0],
\end{aligned} \quad (2)$$

where $AZ_{k,t}^{*,m}$ and $\varepsilon_{k,t}^m$, denote the latent variable and the error term, respectively, with superscript $m$ indicating the model specification based on MHB12. The $\delta$s are unknown parameters.

In Equation (3), we present the model specification based on BIL04:[10]

$$\begin{aligned}
AZ_{k,t}^{*,b} = &\; \gamma_1 \text{PAX}_{k,t} + \sum_i \gamma_{2,i} \text{DIST } i_k + \gamma_3 \text{POP}_{k,t} + \gamma_4 \text{VACATION}_{k,t} + \gamma_5 \text{MAXINC}_{k,t} + \\
& \gamma_6 \text{MININC}_{k,t} + \gamma_7 \text{MAXAZCIT}_{k,t} + \gamma_8 \text{MINAZUCIT}_{k,t} + \gamma_9 \text{ZERAZCIT}_{k,t} + \\
& \gamma_{10} \text{AZSHCON}_{k,t} + \gamma_{11} \text{HUBOTH}_{k,t} + \gamma_{12} \text{HHI}_{k,t} + \gamma_{13} \text{MAXHHI}_{k,t} \times \text{MEDSMA}_{k,t} +
\end{aligned} \quad (3)$$

---

[9] Müller, Hüschelrath and Bilotkach (2012), Table 2.
[10] Boguslaski, Ito and Lee (2004), p. 334, Table V.

$$\gamma_{14}\text{MAXHHI}_{k,t} \times \text{BIG}_{k,t} + \gamma_{15}\text{MINHHI}_{k,t} + \gamma_{16}\text{LCCCOMP}_{k,t} + \varepsilon^b_{k,t}, \text{ AZ}_{k,t} = \mathbb{1}[\text{AZ}^{*,b}_{k,t} > 0],$$

where $\text{AZ}^{*,b}_{k,t}$ and $\varepsilon^b_{k,t}$, denote the latent variable and the error term, respectively, with superscript $m$ indicating the model specification based on BIL04. The $\gamma$s are unknown parameters.

Note that Equations (2) and (3) contain versions similar to those proposed by BIL04 and MHB12. As can be seen from the description of the regressors (Table 1 and Appendix), the metrics used for each variable are not exactly the same as in the original studies because of the availability of information and the configuration of the data sets. Additionally, to proceed with the estimations, our specifications contain additional controls to the set employed by each study.[11] However, as it is an application to a different case, we must emphasize that the desired comparability between the estimates is far from ideal. We proceeded with the application of the methodology, however, to examine not only the level of agreement among the raters but also to assess the strengths and weaknesses of our proposed comparative approach.

We considered the following raters for MHB12 and BIL04:[12]

- JB(NS): original MHB12 estimates for JetBlue, all non-stop entries;
- JB(NS,EXIST): original MHB12 estimates for JetBlue, non-stop entry into existing non-stop markets;
- JB(NS,NEW): original MHB12 estimates for JetBlue, non-stop entry into new markets.
- SW(PER1): original BIL04 estimates for southwest, for entries between 1991 and 2000 (period 1);
- SW(PER2): original BIL04 estimates for southwest, for entries between 1995 and 2000 (period 2).

Additionally, we set the raters for the Azul entry models (the BIL04-like and MHB12-like estimates) as follows (the full empirical results of the estimations of each econometric specification that comprise these raters are available in the Appendix).

- AZ(FULL): all non-stop entries, full sample period;
- AZ(BEF): all non-stop entries, sample period before the Azul-Trip merger.
- AZ(AFT): all non-stop entries, sample period after the Azul-Trip merger.
- AZ(BEF,EXIST): sample period before the Azul-Trip merger; non-stop entry into existing markets;

---

[11] The additional controls are TREND, TREND × DIST, TREND × HUB, TREND × SECND, and TREND × NEW.

[12] See details in Müller, Hüschelrath and Bilotkach (2012), Table 2, and Boguslaski, Ito and Lee (2004), p. 334, Table V.

- AZ(BEF,NEW): sample period before the Azul-Trip merger; non-stop entry into new markets;
- AZ(AFT,EXIST): sample period after the Azul-Trip merger; non-stop entry into existing markets;
- AZ(AFT,NEW): sample period after the Azul-Trip merger; non-stop entry into new markets.

Table 3 presents the results of our empirical methodology for entry pattern comparison between Azul on the one hand, and JetBlue and Southwest on the other. It displays the kappa statistic measures of agreement between the proposed raters; the grayscale reflects the Landis and Koch (1997) intervals for the strength of agreement.

**Table 3 - Comparing Azul with JetBlue and Southwest: Kappa-statistic measure of interrater agreement**

| Airline | JB(NS) | JB(NS,EXIST) | JB(NS,NEW) | SW(PER2) | SW(PER1) |
|---|---|---|---|---|---|
| | MHB12 | MHB12 | BIL04 | MHB12 | BIL04 |
| AZ(BEF) | 0.518*** | 0.167 | 0.300 | 0.209 | 0.149 |
| AZ(BEF,EXIST) | 0.388** | 0.352*** | 0.300* | 0.267* | 0.055 |
| AZ(BEF,NEW) | 0.273 | 0.200 | 0.375* | -0.071 | -0.038 |
| AZ(AFT,EXIST) | -0.035 | 0.210 | 0.070 | -0.064 | -0.042 |
| AZ(AFT,NEW) | -0.113 | 0.020 | -0.125 | -0.202 | -0.073 |
| AZ(AFT) | -0.202 | -0.014 | -0.302* | -0.064 | -0.042 |
| AZ(FULL) | -0.041 | 0.077 | -0.310** | -0.045 | -0.085 |

Landis and Koch (1977) scale for strength of agreement

| below 0.0 | 0.00–0.20 | 0.21–0.40 | 0.41–0.60 | 0.61–0.80 | 0.81–1.00 |
|---|---|---|---|---|---|
| Poor | Slight | Fair | Moderate | Substantial | Almost perfect |

*Notes: Kappa measures calculated by comparing the estimated coefficients of Table 6 and Table 7 (available in the Appendix), with, respectively, the equivalent estimated coefficients of Boguslaski, Ito & Lee (2004, p. 334, Table V) for Southwest Airlines, and Müller, Hüschelrath & Bilotkach (2012, p. 493, Table 2) for JetBlue Airways. "BIL04" denotes Boguslaski, Ito & Lee (2004); "MHB12" denotes Müller, Hüschelrath & Bilotkach (2012). "AZ", "JB", and "SW" denote Azul Airlines, JetBlue Airways, and Southwest Airlines. "FULL", "BEF", and "AFT" denote full sample period, before merger period, and after merger period, respectively; "EXIST" and "NEW" mean existing routes operated by carriers in the market and new routes, respectively. "NS", "NS,NEW", and "NS,EXIST" denote the three non-stop entry cases considered by MHB12; "PER1" and "PER2" denote the two periods considered by BIL04. Approximate normal test statistics and p-values computed using bootstrapped standard errors with 2,000 replications. The null hypothesis is that the agreement between raters is due to chance. P-value representations: \*\*\*p<0.01, \*\* p<0.05, \* p<0.10.*

In Table 3, the agreement by chance hypothesis is rejected at the 1% level in the following cases: AZ(BEF) vs. JB(NS), AZ(BEF,EXIST) vs. JB(NS), and AZ(BEF,EXIST) vs. JB(NS, EXIST). All these results point to a greater consistency between Azul's entry patterns observed before the merger and, more specifically, on existing routes, and JetBlue's entry patterns in all markets and existing markets cases, as estimated by MHB12. For these cases, Table 3 shows fair-to-moderate but statistically significant agreements. However, the agreement among the raters regarding the entry on new routes is weaker, given that the Kappa statistics of AZ(BEF,NEW) vs. JB(BEF,NEW) is equal to 0.375, but statistically significant only at 10%. This result suggests that due to different socioeconomic realities, market entry opportunities in Brazil may also be considerably dissimilar from the US case. In addition, there is a full lack of agreement between the raters in the comparisons of Azul with Southwest Airlines, as no Kappa statistics for these cases are statistically significant at least at 5%. Finally, all estimated Kappa values for comparisons between AZ(AFT) and either of the two airline benchmarks are, in many cases, either close to zero or negative, indicating agreements by chance or even weaker.

We can conclude from the results in Table 3 that there is a relatively strong agreement between the raters regarding Azul's entry patterns before the Trip merger, with those of JetBlue, but there is none with Southwest Airlines' entry decisions. What is more, our results suggest that, after the merger, the Brazilian airline started to move away from both entry benchmarks to develop its own entry patterns in the industry.

To allow for a better visualization of the results in Table 3, Table 4 presents the matrices with detailed results for the case of the two strongest interrater agreements for each airline pair, namely AZ(BEF) vs. JB(NS), and AZ(BEF) vs. SW(PER2). For comparison, at the bottom of the table, we also present the equivalent post-merger results for Azul, namely AZ(AFT) vs. JB(NS), and AZ(AFT) vs. SW(PER2). Note that the agreements between raters are arranged on the diagonals of the matrices, highlighted in gray.

Consistent with the findings of Table 3, we observe in Table 4 that the AZ(BEF) rater presents a stronger agreement with JB(NS) than with SW(PER2). In fact, there are 11 variables (out of 16) in which JB(NS) agrees with AZ(BEF), an agreement rate of 69%. In contrast, there are 10 variables (out of 20) in which the AZ(BEF) and SW(PER2) raters agree—a 50% agreement rate. In both cases, it is clear the rates fall considerably when assessing the agreements with AZ(AFT): from 69% to 18% for JB(NS), and from 50% to 20% for SW(PER2).

**Table 4 - Comparing Azul with JetBlue and Southwest: signs and statistical significance of coefficients
- strongest interrater agreements of each airline pair (before and after merger)**

| Airline | JB(NS) | | | | Airline | SW(PER2) | | | |
|---|---|---|---|---|---|---|---|---|---|
| | Estimate Classification | Negative Significant | Not Significant | Positive Significant | | Estimate Classification | Negative Significant | Not Significant | Positive Significant |
| AZ(BEF) | Negative Significant | DIST SQ, SLOT | | | AZ(BEF, EXIST) | Negative Significant | MAXINC, MININC, ZERAZCIT | | VACATION |
| | Not Significant | MAXHHI, UNEMPL | PAX, LCCCOMP, NONHUB, MAXHHI × NONHUB | HHI, INC | | Not Significant | DIST 1500, HUBOTH, MAXHHI × MEDSMA | DIST 300, DIST 900, MAXAZCIT, HHI | |
| | Positive Significant | | FEE | DIST, NETWEC, EXIST, SECND, POP | | Positive Significant | MAXHHI × BIG | DIST 600, DIST 1200, AZSHCON, MINHHI, LCCCOMP | PAX, POP, MINAZCIT |

| Airline | JB(NS) | | | | Airline | SW(PER2) | | | |
|---|---|---|---|---|---|---|---|---|---|
| | Estimate Classification | Negative Significant | Not Significant | Positive Significant | | Estimate Classification | Negative Significant | Not Significant | Positive Significant |
| AZ(AFT) | Negative Significant | | PAX, NONHUB, FEE | DIST | AZ(BEF, EXIST) | Negative Significant | DIST 1500, ZERAZCIT | DIST 300, DIST 600, DIST 900, DIST 1200 | |
| | Not Significant | DIST SQ, BANKR, SLOT, MAXHHI, UNEMPL | LCCCOMP | HHI, SECND, POP, INC | | Not Significant | MAXINC, MININC, MAXHHI × BIG | | POP |
| | Positive Significant | | MAXHHI × NONHUB | NETWEC, EXIST | | Positive Significant | HUBOTH, MAXHHI × MEDSMA | MAXAZCIT, AZSHCON, HHI, MINHHI, LCCCOMP | PAX, VACATION, MINAZCIT |

*Notes: Signs and statistical significance of the estimated coefficients extracted from Table 6 and Table 7 (available in the Appendix), and the equivalent estimated coefficients of Boguslaski, Ito & Lee (2004, p. 334, Table V) for Southwest Airlines (BIL04), and Müller, Hüschelrath & Bilotkach (2012, p. 493, Table 2) for JetBlue Airways (MHB12). "AZ", "JB", and "SW" denote Azul Airlines, JetBlue Airways, and Southwest Airlines. "BEF" and "AFT" denote the pre-merger period, and the post-merger period, respectively. "NS" denotes the first non-stop entry case considered by MHB12. "EXIST" denotes existing routes already operated by carriers in the market; "PER2" denotes the second period considered by BIL04.*

The interrater agreements displayed in Table 4 indicate that both Azul (before the merger) and JetBlue view factors such as DIST, NETWEC, EXIST, SECND, and POP as attractors of entry, and SLOT as a repellant, suggesting an entry barrier. They also agree on a decreasing marginal effect of distance, as dictated by the common negative DIST SQ. Additionally, both airlines do not seem to consider PAX, LCCCOMP, NONHUB, and MAXHHI × NONHUB in their network planning. In contrast, the interrater analysis suggests that airlines consider different roles for factors such as HHI, MAXHHI, INC, UNEMPL, and FEE. This latter result points to the fact that even the young Azul had important entry pattern particularities when contrasted with the considered low-cost airline benchmarks.

As seen in the bottom matrices of Table 4, these idiosyncrasies are further accentuated after the merger with the regional trip as, in these cases, the interrater agreement becomes much weaker. The most important changes in Azul's entry pattern from the analysis of AZ(BEF) vs. JB(NS) to AZ(AFT) vs. JB(NS) are associated with factors such as DIST (becomes negative), POP (becomes insignificant), NONHUB (becomes negative), and MAXHHI × NONHUB (becomes positive). Other changes are related to PAX, SECND, SLOT, and FEE, but we note that the results of the MHB12-like specification are not consistent with the results of our full empirical model of Equation (1), displayed in Columns (4) and (5) of Table 2.

Again, the results allow us to infer that since the 2012 merger, Azul has intensified the adoption of an idiosyncratic entry strategy to succeed, probably adapting its business model to the specificities of the competitive environment in the Brazilian air transport market.

## 5. Conclusion

This study developed an empirical model of the network construction of Azul Airlines in the Brazilian domestic airline industry from 2008 to 2018. Using discrete-choice models with specifications built upon the previous literature, along with some case-specific covariates, we aimed to identify the market characteristics that influenced the carrier's route entry probabilities across the country in the sample period. We also proposed a novel methodology for comparing our results with the findings of the previous literature on empirical determinants of entry by utilizing the kappa statistic for interrater agreement. Our main contribution lies in the investigation of the effects of the 2012 merger with the regional carrier trip airlines and how this event has affected Azul's route entry decisions. We contrasted the estimated entry patterns before and after the merger with the benchmarks of JetBlue Airways and Southwest Airlines in the US air transport market.

Our empirical results indicate a fair-to-moderate but statistically significant agreement between the proposed entry pattern raters of JetBlue and Azul before the merger event. This finding is consistent with the fact that the carriers were founded by the same businessperson, David Neeleman. We obtained strong statistical evidence of agreement for entries on previously existing routes but not on new ones. We found that both Azul (before the merger) and JetBlue considered factors such as network economies, population size, and flights from a secondary airport as positive entry drivers. In contrast, airport slot restrictions constitute an entry barrier for carriers. We also found no consistency between Azul and Southwest Airlines' entry patterns in the nineties. Finally, our results suggest that after the 2012 merger, Azul has moved away from both airline benchmarks to develop its idiosyncratic entry patterns. It has become more interested in short-haul routes connecting cities with lower populations and more concentrated non-hub airports, with strong elements of regional operations to keep expanding and enjoying monopoly positions across the country at the same time. This strategic revamp has promoted Azul to a unique situation when compared to other airlines in the country and around the world.

Consistent with its profile of incessant search for growth through adaptation and opportunity-seeking, Azul has recently reformulated its business model once again, this time with the introduction of Airbus A320 neo aircraft in the domestic market. The move shows the carrier's motivation to become the biggest airline in Brazil by capturing a higher market stake on the denser routes dominated by its major rivals. Neeleman's recent statement that Azul is interested in buying the LATAM Airlines Group is a clear indication of these ambitions.[13] The emergence of a large carrier with entry patterns fully adapted to the country's reality may represent the strengthening of competition in the industry, at least while route and airport concentrations do not increase considerably.

The results of this study have important implications for the air transport industry. Numerous studies have shown the positive effects of an entry into a new destination on the aviation industry, or even to adjacent sectors such as tourism. By better understanding how airlines choose their new destinations through the findings of this study, regulators and government can assess if an airport needs an investment to serve the demand for a potential entry, leading to better expenditure planning.

However, the study has crucial limitations regarding its scope and methods. Future research should investigate market exit determinants to analyze the whole picture of an airline's network planning strategies over time. This is an important step, as the determinants of exit decisions are usually different from the determinants of entry decisions due to many factors such as switching costs. By accounting for the exit decisions of the post-merger analysis, it would be possible to identify the motivations of the merged entity to drop low-profit routes. In this regard, new studies could consider the airline's sunk cost, such as the investments made at the airport, operational structure and start-up costs, and advertising and switching costs. With a structural model of combined route entry and exit decisions, we believe that future studies could better address issues such as the endogeneity of key regressors in our analysis. We also recommend that future studies compare carriers before and after the COVID-19 pandemic to inspect how network configuration decisions have evolved in the air transport industry across different airline business models.

---

[13] "Brazil's Azul eyeing bid for whole of LATAM Airlines, CEO says," *Reuters*, November 2, 2021. Available at www.reuters.com.

**Appendix**

Below is a full list of variables employed in the econometric models, sorting in ascending alphabetic order.

- AZ is a binary response variable indicating the entry of Azul Airlines on a directional airport-pair in a given year. It is assigned with value one only for the year the carrier starts flight operations on the route, being equal to zero for the other years. AZ(FULL), AZ(BEF), and AZ(AFT) mean, respectively, the variable considering the full sample period (2008-2018), the period before the Azul-Trip merger (2008-2011), and the period after that event (2012-2018). AZ(BEF,EXIST) and AZ(AFT,EXIST) denote Azul's entries on existing routes with respect to the 2007 networks of all carriers, respectively before and after the merger. AZ(BEF,NEW) and AZ(AFT,NEW) denote Azul's entries on new routes with respect to the 2007 networks of all carriers, respectively before and after the merger. Source: ANAC's Air Transport Statistical Database.

- AZSHCON is Azul's market share of passengers, served by connecting flights. Source: ANAC's Air Transport Statistical Database (Cotran - plane change passengers).

- BANKR is a dummy of route presence of bankrupt carrier Avianca in 2018. The carrier filed for reorganization in late 2018 and ceased operations in 2019. Source: ANAC's Air Transport Statistical Database.

- BIG is a dummy variable taking the value 1 if the more concentrated endpoint airport has one million passengers or more. Source: ANAC's Air Transport Statistical Database.

- DIST 300, DIST 600, DIST 900, DIST 1200, DIST 1500 are dummy variables that take the value one if the great circle distance in miles of the airport-pair is within the intervals [300, 600), [600, 900), [900, 1200), [1200, 1500), [1500, 3000), respectively, and zero otherwise. The base case of the dummies comprises distances in the interval [100,300). Routes with distances below 100 miles and above 3000 miles are discarded. Source: ANAC's Air Transport Statistical Database.

- DIST is the great circle distance of the airport-pair in 100 miles. Source: ANAC's Air Transport Statistical Database.

- DIST SQ is the square of DIST. Source: ANAC's Air Transport Statistical Database.

- EXIST is a dummy variable to account for the operation of flights in routes that were already operated in 2007. Source: ANAC's Air Transport Statistical Database.

- FEE is the geometric mean of the landing fees on the endpoint airports (inflation-adjusted local currency values, in logarithm). As most medium- and small-sized airports in the country charge very similar fees due to same airport categorization, this variable is computed only for routes including major hub airports, as defined by variable "BIG". Source: ANAC's airport regulations.

- FSCMAJ is a dummy variable to account for the route presence of the major full-service carrier LATAM Airlines in 2007. Source: ANAC's Air Transport Statistical Database.

- HHI is the Herfindahl-Hirschman index of airport pair concentration, considering the revenue passengers of carriers. This figure is computed for 2007. Source: ANAC's Air Transport Statistical Database.

- HUB is a dummy variable taking the value one if at least one of the endpoint airports is Azul's hub, as stated in "Azul S.A. Form 20-F." Filed with the United States Securities and Exchange Commission, April 30, 2021, available at ri.voeazul.com.br. Source: ANAC's Air Transport Statistical Database.

- HUBOTH is the maximum share of domestic connecting passengers between the endpoint airports of the airport pair, across Azul's rivals LATAM, Gol and Avianca. Source: ANAC's Air Transport Statistical Database.

- INC is a proxy for the mean income of consumers. It is equal to the per capita gross domestic product's geometric mean between the origin and destination cities (inflation-adjusted local currency values, in logarithm). To compute this variable, we consider the city's mesoregion (grouping of nearby cities). Source: IBGE.

- LCCCOMP is a dummy variable to account for the route presence of low-cost carriers (LCC) Gol, Webjet, and BRA in 2007. Source: ANAC's Air Transport Statistical Database.

- LCCMAJ is a dummy variable to account for the route presence of the major low-cost carrier Gol Airlines in 2007. Source: ANAC's Air Transport Statistical Database.

- MAXAZCIT is equal to the maximum of the number of airports Azul serves from the endpoint airports. Source: ANAC's Air Transport Statistical Database.

- MAXHHI is the maximum Herfindahl-Hirschman index of concentration between the endpoint airports of the airport pair, considering the revenue passengers of carriers. This figure is computed for 2007. Source: ANAC's Air Transport Statistical Database.

- MAXHHI × BIG is an interaction variable expressing the multiplication of MAXHHI and BIG.

- MAXHHI × MEDSMA is an interaction variable expressing the multiplication of MAXHHI and MEDSMA.

- MAXHHI × NONHUB is an interaction variable expressing the multiplication of MAXHHI and NONHUB.

- MAXINC is the maximum per capita gross domestic product between the origin and destination airports' cities (inflation-adjusted local currency values, in logarithm). To compute this variable, we consider the city's mesoregion (grouping of nearby cities). Source: IBGE.

- MEDSMA is a dummy variable taking the value one if the less concentrated endpoint airport has less than one million passengers. Source: ANAC's Air Transport Statistical Database.

- MINAZCIT is equal to the minimum of the number of airports Azul serves from the endpoint airports. Source: ANAC's Air Transport Statistical Database.

- MINHHI is the minimum Herfindahl-Hirschman index of concentration between the endpoint airports of the airport pair, considering the revenue passengers of carriers. This figure is computed for 2007. Source: ANAC's Air Transport Statistical Database.

- MININC is the minimum per capita gross domestic product between the origin and destination airports' cities (inflation-adjusted local currency values, in logarithm). To compute this variable, we consider the city's mesoregion (grouping of nearby cities). Source: IBGE.

- NETWEC is equal to the sum of the number of routes Azul server from the endpoint airports, a proxy for the number of potential new connection routes. Source: ANAC's Air Transport Statistical Database.

- NEW is a dummy variable to account for the operation of flights in routes that were not operated in 2007. Source: ANAC's Air Transport Statistical Database.

- NONHUB is a dummy variable taking the value one if at least one of the endpoint airports has a national passenger share below 0.25%. Source: ANAC's Air Transport Statistical Database.

- PAX is the total number of revenue passengers carried by all airlines on the airport-pair in 2007 (in logarithm). Source: ANAC's Air Transport Statistical Database.

- POP is the geometric mean of the population of the origin and destination cities in ten thousands (in logarithm). To compute this variable, we consider the city's mesoregion (grouping of nearby cities). Source: IBGE.

- REGSMA is a dummy variable to account for the route presence of small regional carriers in 2007. Source: ANAC's Air Transport Statistical Database.

- SECND is a dummy variable to indicate if one of the endpoints is the secondary airport São Paulo/Campinas Airport (VCP), and zero otherwise. Source: ANAC's Air Transport Statistical Database.

- SLOT is a dummy variable to indicate if one of the endpoints is a slot-constrained airport, and zero otherwise. In the sample period the slot-constrained airports are São Paulo/Congonhas (CGH), São Paulo/Guarulhos (GRU), Rio de Janeiro/Santos Dumont (SDU), and Belo Horizonte/Pampulha (PLU). Source: ANAC' Slot Coordination webpage (www.anac.gov.br/en/air-services/slot-coordination), July 2, 2017, retrieved from web.archive.org.

- TREND is a time trend variable, equal to 1, 2, ..., T, where T is the total number of sample years (eleven).

- TREND × DIST is an interaction variable expressing the multiplication of TREND and DIST.

- TREND × HUB is an interaction variable expressing the multiplication of TREND and HUB.

- TREND × NEW is an interaction variable expressing the multiplication of TREND and NEW.

- TREND × SECND is an interaction variable expressing the multiplication of TREND and SECND.

- UNEMPL is the geometric mean of a proxy for the formal unemployment rate of the origin and destination cities. The formal employment rate is computed as the number of workers with active employment link on December 31 of each year over the working age population (people aged 15 to 64). The formal unemployment rate proxy is one minus the formal employment rate. To compute this variable, we consider the city's mesoregion (grouping of nearby cities). Sources: Annual Social Information Report (RAIS Microdata), Ministry of Labor and Social Security (Brazil) and IBGE's 2010 Population Census.

- VACATION is is the geometric mean of a proxy for tourism intensity of the origin and destination cities' states. Tourism intensity is defined as the percentage of tourism-related activities' gross revenues with respect to the gross domestic product of the state. Sources: IBGE's Annual Survey of Services, and Gross Domestic Product of Municipalities Publication.

- ZERAZCIT is a dummy variable taking the value one if Azul serves neither of the endpoint airports. Source: ANAC's Air Transport Statistical Database.

**Table 5 - Correlation matrix of the main model variables**

| | | Person correlation coefficient | | | | | | | | | | | | | | | | | | | | | |
|---|---|---|---|---|---|---|---|---|---|---|---|---|---|---|---|---|---|---|---|---|---|---|---|
| | Variable | (1) | (2) | (3) | (4) | (5) | (6) | (7) | (8) | (9) | (10) | (11) | (12) | (13) | (14) | (15) | (16) | (17) | (18) | (19) | (20) | (21) | (22) | (23) |
| (1) | AZ(FULL) | 1 | | | | | | | | | | | | | | | | | | | | | | |
| (2) | PAX | 0.19 | 1 | | | | | | | | | | | | | | | | | | | | | |
| (3) | DIST | -0.03 | -0.07 | 1 | | | | | | | | | | | | | | | | | | | | |
| (4) | POP | 0.05 | 0.14 | -0.16 | 1 | | | | | | | | | | | | | | | | | | | |
| (5) | INC | 0.04 | 0.07 | -0.28 | 0.39 | 1 | | | | | | | | | | | | | | | | | | |
| (6) | UNEMPL | -0.05 | -0.12 | 0.22 | -0.51 | -0.85 | 1 | | | | | | | | | | | | | | | | | |
| (7) | VACATION | 0.03 | 0.04 | -0.18 | 0.30 | 0.20 | -0.22 | 1 | | | | | | | | | | | | | | | | |
| (8) | SECND | 0.04 | 0.06 | -0.03 | 0.18 | 0.08 | -0.09 | 0.04 | 1 | | | | | | | | | | | | | | | |
| (9) | SLOT | 0.05 | 0.16 | -0.05 | 0.29 | 0.13 | -0.17 | 0.09 | -0.01 | 1 | | | | | | | | | | | | | | |
| (10) | FEE | 0.16 | 0.47 | 0.00 | 0.15 | 0.08 | -0.14 | 0.05 | 0.14 | 0.10 | 1 | | | | | | | | | | | | | |
| (11) | NETWEC | 0.14 | 0.21 | -0.02 | 0.30 | 0.20 | -0.27 | 0.15 | 0.51 | 0.19 | 0.30 | 1 | | | | | | | | | | | | |
| (12) | ZERAZCIT | -0.05 | -0.09 | -0.01 | -0.18 | -0.20 | 0.22 | -0.24 | -0.10 | -0.11 | -0.12 | -0.49 | 1 | | | | | | | | | | | |
| (13) | AZSHCON | 0.26 | 0.29 | -0.04 | 0.09 | 0.05 | -0.07 | 0.04 | 0.18 | 0.04 | 0.32 | 0.30 | -0.08 | 1 | | | | | | | | | | |
| (14) | HUBOTH | 0.05 | 0.12 | 0.05 | 0.20 | 0.12 | -0.25 | 0.04 | 0.02 | 0.17 | 0.14 | 0.31 | -0.41 | 0.06 | 1 | | | | | | | | | |
| (15) | NONHUB | -0.08 | -0.20 | 0.00 | -0.21 | -0.12 | 0.24 | -0.06 | -0.11 | -0.19 | -0.24 | -0.40 | 0.32 | -0.11 | -0.44 | 1 | | | | | | | | |
| (16) | HHI | 0.16 | 0.78 | -0.06 | 0.12 | 0.06 | -0.10 | 0.03 | 0.06 | 0.13 | 0.37 | 0.19 | -0.09 | 0.27 | 0.12 | -0.18 | 1 | | | | | | | |
| (17) | MAXHHI | 0.00 | -0.01 | 0.10 | -0.08 | -0.09 | 0.16 | -0.09 | 0.01 | -0.01 | -0.02 | 0.03 | -0.14 | -0.01 | 0.06 | -0.01 | 0.01 | 1 | | | | | | |
| (18) | FSCMAJ | 0.15 | 0.76 | -0.04 | 0.13 | 0.07 | -0.11 | 0.04 | 0.08 | 0.12 | 0.49 | 0.19 | -0.07 | 0.25 | 0.10 | -0.19 | 0.49 | -0.02 | 1 | | | | | |
| (19) | LCCMAJ | 0.13 | 0.71 | -0.05 | 0.12 | 0.07 | -0.10 | 0.04 | 0.08 | 0.12 | 0.42 | 0.17 | -0.06 | 0.23 | 0.09 | -0.15 | 0.39 | -0.02 | 0.70 | 1 | | | | |
| (20) | LCCCOMP | 0.16 | 0.78 | -0.05 | 0.14 | 0.07 | -0.11 | 0.05 | 0.08 | 0.13 | 0.46 | 0.20 | -0.08 | 0.27 | 0.12 | -0.18 | 0.52 | -0.02 | 0.69 | 0.77 | 1 | | | |
| (21) | BANKR | 0.01 | 0.14 | -0.01 | 0.03 | 0.02 | -0.02 | 0.02 | 0.00 | 0.03 | 0.11 | 0.04 | -0.02 | 0.04 | 0.01 | -0.03 | 0.07 | 0.00 | 0.13 | 0.15 | 0.13 | 1 | | |
| (22) | REGSMA | 0.11 | 0.49 | -0.10 | 0.06 | 0.04 | -0.05 | 0.00 | 0.05 | 0.09 | 0.12 | 0.11 | -0.05 | 0.19 | 0.07 | -0.09 | 0.50 | 0.03 | 0.19 | 0.22 | 0.24 | 0.03 | 1 | |
| (23) | NEW | 0.26 | -0.01 | -0.04 | 0.07 | 0.05 | -0.07 | 0.02 | 0.07 | 0.07 | 0.17 | 0.16 | -0.06 | 0.23 | 0.08 | -0.10 | 0.06 | 0.00 | -0.01 | 0.00 | -0.01 | 0.01 | -0.01 | 1 |

## Table 6 - Estimation results: Azul entries - model specification based on BIL04

| | (1) AZ(FULL) | (2) AZ(FULL) | (3) AZ(BEF) | (4) AZ(AFT) | (5) AZ(BEF,EXIST) | (6) AZ(AFT,EXIST) | (7) AZ(BEF,NEW) | (8) AZ(AFT,NEW) |
|---|---|---|---|---|---|---|---|---|
| PAX | 0.0202*** | 0.0699*** | 0.0457*** | 0.0814*** | 0.0456*** | 0.0443*** | – | – |
| DIST 300 | -0.2395*** | -0.2291*** | -0.0352 | -0.2641*** | -0.0913 | -0.3170*** | 0.3248 | -0.1367** |
| DIST 600 | -0.4526*** | -0.4419*** | 0.3641** | -0.5858*** | 0.4000** | -0.6937*** | 0.4740* | -0.3075*** |
| DIST 900 | -0.5831*** | -0.4863*** | 0.2113 | -0.6691*** | 0.2266 | -0.7470*** | 0.4249 | -0.4067*** |
| DIST 1200 | -0.6111*** | -0.4861*** | 0.6369*** | -0.7632*** | 0.5970** | -0.9595*** | 0.8276*** | -0.3336*** |
| DIST 1500 | -0.9967*** | -0.7909*** | 0.0804 | -1.1019*** | 0.1566 | -1.2282*** | – | -0.7489*** |
| POP | 0.0716*** | -0.0079 | 0.3287*** | -0.0349 | 0.3521*** | -0.0005 | 0.2548** | -0.0540** |
| VACATION | -0.4663** | 1.0346*** | -1.9936** | 1.2132*** | -2.3621** | 0.9897*** | 0.3875 | 0.6483* |
| MAXINC | 0.0463 | -0.0093 | -0.2713** | -0.0413 | -0.3175*** | -0.0105 | 0.1719 | -0.0291 |
| MININC | 0.0221 | -0.0104 | -0.2603*** | -0.0058 | -0.3150*** | -0.0482 | 0.2267 | 0.0881** |
| MAXAZCIT | 0.0084*** | 0.0329*** | 0.0160 | 0.0285*** | 0.0093 | 0.0292*** | 0.0399 | 0.0353*** |
| MINAZCIT | 0.0706*** | 0.0521*** | 0.1196*** | 0.0536*** | 0.1197*** | 0.0602*** | 0.0556* | 0.0494*** |
| ZERAZCIT | -0.5150*** | -0.7402*** | -1.5411*** | -0.1622*** | -1.5470*** | -0.1332*** | – | -0.1836*** |
| AZSHCON | 0.7968*** | 0.5904*** | 0.5240*** | 0.5704*** | 0.4856*** | 0.4888*** | 0.6407** | 1.1087*** |
| HUBOTH | 0.4829*** | 0.1351*** | 0.1313 | 0.2811*** | 0.1198 | 0.4250*** | 0.2962 | 0.3258*** |
| HHI | 0.4855*** | 0.4362*** | 0.2330** | 0.4687*** | 0.2260* | 0.8809*** | – | – |
| MAXHHI × MEDSMA | 0.0477*** | 0.0507*** | 0.0370 | 0.0505*** | 0.0415* | 0.0426*** | 0.0464 | 0.0825*** |
| MAXHHI × BIG | 0.0727*** | 0.0286** | 0.2267*** | -0.0125 | 0.2537*** | 0.0219 | 0.1335** | 0.0018 |
| MINHHI | 0.5800*** | 0.7189*** | 1.1322*** | 0.6979*** | 1.2857*** | 0.9307*** | 0.4937** | 0.2747*** |
| LCCCOMP | -0.0251 | 0.2731*** | 0.3272** | 0.2588*** | 0.3861*** | 0.3127*** | – | – |
| TREND | | -0.1293*** | -0.0521 | -0.1916*** | 0.0084 | -0.1475*** | 0.0569 | -0.0218* |
| TREND × DIST | | -0.0006 | -0.0326*** | 0.0028** | -0.0335*** | 0.0035** | -0.0130*** | 0.0001 |
| TREND × HUB | | -0.0939*** | 0.1245*** | -0.0810*** | 0.0988*** | -0.0826*** | 0.1847*** | -0.0549*** |
| TREND × SECND | | -0.1440*** | -0.0580 | -0.1318*** | -0.0418 | -0.2125*** | -0.1880 | -0.1263*** |
| TREND × NEW | | 0.2414*** | 0.2560*** | 0.2527*** | – | – | – | – |
| Estimator | PROBIT | PROBIT | PROBIT | PROBIT | PROBIT | PROBIT | PROBIT | PROBIT |
| Airport-Pair Clusters | 95,698 | 95,698 | 95,698 | 95,698 | 95,698 | 95,698 | 95,698 | 95,698 |
| Log likelihood Statistic | -7,887 | -6,051 | -590 | -5,135 | -513 | -3,223 | -155 | -4,022 |
| Pseudo R2 Statistic | 0.4210 | 0.5558 | 0.6473 | 0.5582 | 0.6485 | 0.5254 | 0.4735 | 0.3275 |
| AIC Statistic | 15,816 | 12,153 | 1,232 | 10,322 | 1,076 | 6,496 | 350 | 8,088 |
| BIC Statistic | 16,065 | 12,462 | 1,515 | 10,619 | 1,347 | 6,782 | 567 | 8,340 |
| Nr Observations | 1,052,678 | 1,052,678 | 382,792 | 669,886 | 382,792 | 669,886 | 382,792 | 669,886 |

*Notes: Estimation results produced by the probit model, denoted as "PROBIT". Standard errors of estimates allow for intragroup correlation (clustered sandwich estimator), using airport pairs as clusters. "–" denotes that the variable is dropped. "AZ" denotes the probability of Azul's route entry. "FULL", "BEF", and "AFT" denote full sample period, before merger period, and after merger period, respectively. "EXIST" and "NEW" mean existing routes operated by carriers and new routes, respectively. P-value representations: \*\*\*p<0.01, \*\* p<0.05, \* p<0.10.*

<div align="center">**Table 7 - Estimation results: Azul entries - model specification based on MHB12**</div>

| | (1) AZ(FULL) | (2) AZ(FULL) | (3) AZ(BEF) | (4) AZ(AFT) | (5) AZ(BEF,EXIST) | (6) AZ(AFT,EXIST) | (7) AZ(BEF,NEW) | (8) AZ(AFT,NEW) |
|---|---|---|---|---|---|---|---|---|
| DIST | -0.0701*** | -0.0268** | 0.1917*** | -0.0893*** | 0.1026*** | -0.2070*** | 0.1588*** | -0.1207*** |
| DIST SQ | 0.0011*** | 0.0004 | -0.0075*** | 0.0004 | -0.0036** | 0.0034*** | -0.0071** | 0.0014*** |
| PAX | -0.0521*** | -0.0168*** | 0.0039 | -0.0281*** | 0.0628*** | 0.0448*** | – | – |
| HHI | 0.4563*** | -0.0150 | -0.0368 | 0.0003 | 0.4052*** | 1.0801*** | – | – |
| LCCCOMP | -0.1121** | 0.0219 | -0.0452 | 0.0032 | 0.3242** | 0.3958*** | – | – |
| BANKR | -0.5377*** | -0.2761 | – | 0.2274 | – | -0.1313 | – | – |
| NETWEC | 0.0288*** | 0.0424*** | 0.0512*** | 0.0227*** | 0.0464*** | 0.0375*** | 0.0645*** | 0.0463*** |
| EXIST | 1.2470*** | 2.2667*** | 2.1207*** | 2.8875*** | – | – | – | – |
| SECND | -0.8285*** | 0.9927*** | 1.6889*** | -0.4919* | 1.7611*** | -0.1735 | 0.6211* | -0.8424*** |
| SLOT | 0.0825** | -0.1973*** | -0.8468*** | -0.0662 | -0.9121*** | -0.1524** | -0.0235 | 0.0087 |
| MAXHHI | -0.1497** | -0.0870 | -0.6397* | -0.1587 | -0.4185** | -0.2902*** | -0.1500 | 0.0752 |
| NONHUB | -0.5483*** | -0.2159*** | -0.4241 | -0.2929*** | -0.7059*** | -0.6454*** | -0.4629 | -0.2808*** |
| MAXHHI × NONHUB | 0.3749*** | 0.2777*** | 0.3300 | 0.3345*** | 0.4246* | 0.4896*** | 0.3202 | 0.1511* |
| FEE | 0.1444*** | -0.0494*** | 0.2310*** | -0.0824*** | 0.3993*** | 0.0407 | 0.2768*** | 0.1399*** |
| POP | 0.0055 | -0.0209 | 0.4820*** | -0.0196 | 0.4504*** | 0.0350 | 0.2872*** | -0.0628*** |
| INC | -0.0779 | 0.0483 | -0.4640* | 0.0073 | -0.7090*** | -0.0348 | 0.6542 | -0.1022 |
| UNEMPL | -0.0106*** | -0.0006 | 0.0046 | 0.0030 | -0.0115 | 0.0014 | 0.0139 | -0.0075** |
| TREND | | -0.0616*** | 0.3478*** | -0.3001*** | 0.3731*** | -0.1825*** | 0.0024 | -0.0363*** |
| TREND × DIST | | -0.0019* | -0.0346*** | 0.0072*** | -0.0304*** | 0.0115*** | -0.0127 | 0.0052*** |
| TREND × HUB | | -0.1165*** | 0.1013*** | -0.0617*** | 0.0619 | -0.0905*** | 0.1430** | -0.0777*** |
| TREND × SECND | | -0.2831*** | -0.6967*** | -0.0456 | -0.6709*** | -0.1793** | -0.4654*** | -0.0195 |
| TREND × NEW | | 0.3330*** | 0.5645*** | 0.3913*** | – | – | – | – |
| Estimator | PROBIT | PROBIT | PROBIT | PROBIT | PROBIT | PROBIT | PROBIT | PROBIT |
| Airport-Pair Clusters | 95,698 | 95,698 | 95,698 | 95,698 | 95,698 | 95,698 | 95,698 | 95,698 |
| Log likelihood Statistic | -8,199 | -5,416 | -584 | -4,289 | -564 | -3,350 | -167 | -4,548 |
| Pseudo R2 Statistic | 0.3981 | 0.6024 | 0.6511 | 0.6310 | 0.6133 | 0.5066 | 0.4336 | 0.2397 |
| AIC Statistic | 16,435 | 10,879 | 1,212 | 8,624 | 1,168 | 6,743 | 368 | 9,129 |
| BIC Statistic | 16,648 | 11,152 | 1,451 | 8,887 | 1,385 | 6,983 | 552 | 9,323 |
| Nr Observations | 1,052,678 | 1,052,678 | 382,792 | 669,886 | 382,792 | 669,886 | 382,792 | 669,886 |

*Notes: Estimation results produced by the probit model, denoted as "PROBIT". Standard errors of estimates allow for intragroup correlation (clustered sandwich estimator), using airport pairs as clusters. "–" denotes that the variable is dropped. "AZ" denotes the probability of Azul's route entry. "FULL", "BEF", and "AFT" denote full sample period, before merger period, and after merger period, respectively. "EXIST" and "NEW" mean existing routes operated by carriers and new routes, respectively. P-value representations: \*\*\*p<0.01, \*\* p<0.05, \* p<0.10..*